\documentclass[twocolumn,showpacs,preprintnumbers,amsmath,amssymb,aps,pre,floatfix]{revtex4}

\usepackage{graphicx}
\usepackage{dcolumn}
\usepackage{bm}
\begin{document}

\preprint{LA-UR 09-01633}

\title{An improved model for the transit entropy of monatomic liquids}

\author{Duane C. Wallace}
\author{Eric D. Chisolm}
\author{Nicolas Bock}

\affiliation{Theoretical Division, Los Alamos National Laboratory, 
Los Alamos, New Mexico 87545, USA}

\date{\today}

\begin{abstract}
In the original formulation of vibration-transit (V-T) theory for monatomic liquid
dynamics, the transit contribution to entropy was taken to be a
universal constant, calibrated to the constant-volume entropy of
melting.  This model suffers two deficiencies: (a) it does not account
for experimental entropy differences of $\pm2\%$ among elemental
liquids, and (b) it implies a value of zero for the transit
contribution to internal energy.  The purpose of this paper is to
correct these deficiencies.  To this end, the V-T equation for entropy
is fitted to an overall accuracy of $\pm0.1\%$ to the available
experimental high temperature entropy data for elemental liquids.  The
theory contains two nuclear motion contributions: (a) the dominant
vibrational contribution $S_{vib}(T/\theta_0)$, where $T$ is
temperature and $\theta_0$ is the vibrational characteristic
temperature, and (b) the transit contribution $S_{tr}(T/\theta_{tr})$,
where $\theta_{tr}$ is a scaling temperature for each liquid.  The
appearance of a common functional form of $S_{tr}$ for all the liquids
studied is a property of the experimental data, when analyzed via the
V-T formula.  The resulting $S_{tr}$ implies the correct transit
contribution to internal energy.  The theoretical entropy of melting
is derived, in a single formula applying to normal and anomalous
melting alike.  An \textit{ab initio} calculation of $\theta_0$, based
on density functional theory, is reported for liquid Na and Cu.
Comparison of these calculations with the above analysis of
experimental entropy data provides verification of V-T theory.
In view of the present results, techniques currently being applied in
\textit{ab initio} simulations of liquid properties can be employed to
advantage in the further testing and development of V-T theory.
\end{abstract}

\maketitle

\section{Introduction}

We are interested in the theoretical description of the motion of
nuclei (or atoms) in real monatomic liquids.  For many years, such
descriptions have been available for gases and crystals, consisting in
each case of an approximate but tractable ``zeroth order''
Hamiltonian, plus complicated but small corrections.  Zeroth order for
a gas is free particle motion, from Boltzmann \cite{B_WB72a}, and the
correction is potential energy.  Zeroth order for a crystal is
harmonic vibrational motion \cite{BvK_PZ12a,BH_54a}, and the major
correction is anharmonicity.  These theories are extremely valuable,
as the zeroth order Hamiltonian provides a complete orthogonal basis
set for the nuclear motion.  Hence for any physically meaningful
problem, the motion can be analyzed and statistical mechanics can be
constructed in terms of the basis set.  These theories account for
equilibrium and nonequilibrium properties of gases and crystals to an
accuracy on the order of experimental accuracy.

Our proposal for this kind of theory for monatomic liquids is
vibration-transit (V-T) theory \cite{W_PRE97b}.  The key postulate is that
the many-body potential energy surface is overwhelmingly dominated by
intersecting macroscopically equivalent random valleys.  The zeroth
order Hamiltonian expresses normal mode vibrational motion in a single
random valley harmonically extended to infinity.  ``Macroscopic
equivalence'' means that this Hamiltonian for any random valley gives
the same statistical averages in the thermodynamic limit
\cite{macro_equiv}.  The motion of nuclei is then composed of two
parts: brief periods of vibration in one random valley, interspersed
with transits which carry the system between neighboring random
valleys.  Transits, each involving a small local group of nuclei,
proceed at a high rate throughout the liquid.  For calculation of the
partition function, the effect of transits is to correct the potential
surface for the valley-valley intersections \cite{W_PRE97b,W_02a}.
For nonequilibrium calculations as, e.g., of time correlation
functions, the same transits provide the diffusive jumps of the nuclei
\cite{W_PRE97b,DCW_PRE08a}.  The vibrational motion is tractable, and
is calibrated from potential properties of a single random valley.
The vibrational contribution to a thermodynamic function is around
$90\%$ of the total \cite{W_PRE97b,W_02a}.  The transit motion is
complicated, but its contribution to a thermodynamic function is only
around $10\%$ \cite{W_PRE97b,W_02a}.  This paper is mainly concerned
with the vibrational and transit contributions to the entropy of
monatomic liquids.

In the original formulation of V-T theory, transits are accounted for
only insofar as they give the liquid access to all the random valleys.
This property is modeled by multiplying the single random valley
partition function by a universal number, calibrated from entropy of
melting data \cite{W_PRE97b}.  For $S_{tr}$, the transit contribution to
entropy, this yields $S_{tr} = 0.8Nk_B$, the same for every
monatomic liquid.  Theory for the total entropy agrees with
high-temperature experimental data for normal melting elements to
within $\pm0.2Nk_B$ (\cite{W_PRE97b}, Fig.~2).  Since $0.2Nk_B$ is
approximately $2\%$ of the total entropy for the elemental liquids,
the original V-T formulation is quite satisfactory for such a simple
model.  Nevertheless, the original formulation suffers two
deficiencies which we wish to correct here.
\begin{enumerate}
\item Since $S_{tr}$ is a universal constant, it does not account for
      the different behaviors of individual liquids.  However, these
      differences largely account for the scatter which results in
      theoretical errors of up to $2\%$ in the entropy.
\item Since $S_{tr}$ does not depend on $V$ or $T$, the consistent
      contribution to all \emph{other} thermodynamic functions is zero.  But
      we now know that the transit contribution to internal energy at melt
      is around $10\%$, and this energy must be included to get a good
      theoretical value of the melting temperature $T_m$.
\end{enumerate}
To correct these deficiencies, we must account explicitly for the $V$
and $T$ dependences of $S_{tr}$.

To this end, in Sec.~\ref{Sanal} we extend the original analysis of
experimental high-temperature entropy data to normal and
anomalous melting elemental liquids.  We find that all the data for
$S_{tr}(T)$ at constant volume can be fitted to a single curve,
providing a scaling formula for the $V$ and $T$ dependences of
$S_{tr}$.  The fitting yields values for each liquid of the
vibrational characteristic temperature $\theta_0$ and of a new transit
characteristic temperature $\theta_{tr}$.  The original ``universal
entropy constant'' has a role in the present formulation, where it is
expected to vary weakly with $V$, and also to vary weakly among the
elemental liquids.

In liquids, as in all condensed matter systems, the potential energy
that governs the nuclear motion is the electronic ground state energy
as a function of nuclear positions \cite{BH_54a,W_02a}.  Hence the
parameters in the liquid dynamics Hamiltonian can be calculated from
electronic structure theory.  Over the years, pseudopotential
perturbation theory for nearly-free-electron (NFE) liquids has been
extremely useful in the study of liquid dynamics for real liquids
\cite{H_66a,A_PL66a,F_72a,AS_SSP78a,M_90a}.  This is the basis of a
series of tests of V-T theory for liquid Na
\cite{DCW_PRE08a,WC_PRE99a,CW_PRE99a,CCW_PRE01a,DW_PRE07a}.  Extension
of this principle beyond NFE liquids, by means of density functional
theory (DFT), is the subject of Sec.~\ref{testingVT}.  First, V-T
theory for the entropy of melting is reviewed, and a single equation
covering normal and anomalous melting is derived.  Results from a new
method \cite{BPCDWHL_08a,HBPLDCW_un09a}, using DFT to calculate the
vibrational parameters, are then applied to test the V-T theory of
entropy for Na and Cu.  \textit{Ab initio} techniques are currently
being applied to a wide range of liquid dynamics studies, and their
potential in testing and developing V-T theory is noted.  In
Sec.~\ref{concl}, broader application of the present reformulation of
the transit entropy is described.  The verification of V-T theory
provided by the present \textit{ab initio} test is also discussed.

\section{Analysis of experimental entropy data}
\label{Sanal}

In V-T theory, the liquid entropy is given by
\begin{equation}
S(V,T) = S_{vib}(V,T) + S_{tr}(V,T) + S_{el}(V,T).
\label{Sdef}
\end{equation}
$S_{vib}$ describes the nuclear motion in a single random valley
harmonically extended to infinity.  In classical statistical
mechanics,
\begin{equation}
S_{vib}(V,T) = 3Nk_B\left\{\ln\left[T/\theta_0(V)\right] + 1\right\}.
\label{classicalSvib}
\end{equation}
The characteristic temperature $\theta_0$ is given by
\begin{equation}
\ln(k_B\theta_0) = \langle \ln(\hbar\omega)\rangle,
\label{Theta0def}
\end{equation}
where $\langle \ldots \rangle$ is the average over the vibrational
normal mode frequencies $\omega$.  The quantum corrections to
Eq.~(\ref{classicalSvib}) are straightforward \cite{W_02a}, and are
negligible in the present analysis.  $S_{tr}$ represents the transit
motion of the nuclei.  $S_{el}$ represents the excitation of electrons
from their ground state with nuclear positions fixed at a random
structure.  Two small contributions, neglected here, express
anharmonicity of the vibrational motion and the interaction between
nuclear motion and electronic excitations (\cite{W_02a}, Sec.~4).

The experimental data are at ambient pressure, where the volume increases
with temperature.  It is most helpful to remove the volume dependence
of the experimental data, by correcting $S_{expt}(V,T)$ to
$S_{expt}(V_m,T)$, where $V_m$ is the fixed volume of the liquid at
melt.  At fixed volume, the parameters of our analysis are simply
constants.  With $(\partial S / \partial V)_T = \beta B_T$, where
$\beta$ is the thermal expansion coefficient and $B_T$ is the
isothermal bulk modulus, the correction to second order is
\begin{equation} 
S_{expt}(V_m,T) = S_{expt}(V,T) + \eta V \beta B_T + 
\frac{1}{2} \eta^2 V^2 \left(\frac{\partial (\beta B_T)}{\partial V}\right)_T,
\label{volumefix}
\end{equation}
where $\eta = (V_m - V)/V$.  For a given liquid the analysis requires
highly accurate experimental data for the entropy and its first volume
correction, at a significant range of temperatures above $T_m$.  The
liquids satisfying this requirement are ten NFE metals.  These are
listed in Table~\ref{maintab}, along with $T_m$, the highest
temperature of our analysis $T_h$, and references for the
experimental data.

\begin{table*}
\caption{Results of the high-temperature entropy analysis for ten
liquids.  $T_m$ is the melting temperature and $T_h$ is the highest
temperature of the analysis.  Experimental data are from the
references cited.  Sn and Ga are anomalous melters \cite{W_PRSL91a}.
$S_{nuc}$ is given by Eq.~(\ref{Snucdef}), and $\langle \delta
S_{expt} \rangle$ is the estimated mean error in the high-$T$ entropy
data after correction to $V=V_m$ (see Appendix \ref{Strerrors}).}
\label{maintab}
\begin{ruledtabular}
\begin{tabular}{cccccccc}
Element & $T_m$ (K) & $T_h$ (K) & References & $\theta_0$ (K) & $\theta_{tr}$ (K) &  $S_{nuc}(T_m)$ ($Nk_B$) & $\langle \delta S_{expt} \rangle$ ($Nk_B$) \\
\hline
Na & $371.0$ & $1100$ & \cite{G_HT66a,YTVPT_HT66a,SC_PRB85a,HDHGKW_73a}        &  $97.6$ & $570$ &  $7.725$ & $0.07$ \\
K  & $336.4$ & $1040$ & \cite{YTVPT_HT66a,SYFSM_85a,SC_PRB85a,CDDFMS_JPCRD85a} &  $58.0$ & $570$ &  $8.986$ & $0.11$ \\
Rb & $312.6$ &  $900$ & \cite{SC_PRB85a,VKSAR_77a,CDDFMS_JPCRD85a}             &  $35.8$ & $530$ & $10.183$ & $0.12$ \\
Cs & $301.6$ &  $948$ & \cite{SC_PRB85a,VKSAR_77a,CDDFMS_JPCRD85a}             &  $26.0$ & $540$ & $11.032$ & $0.11$ \\
Al & $933.5$ & $1400$ & \cite{B_83a,M_52a,CDDFMS_JPCRD85a,TSTS_TJIM82a}        & $198.0$ & $980$ &  $8.451$ & $0.09$ \\
Pb & $600.6$ & $1023$ & \cite{GM_SPA66a,B_83a,M_52a,HDHGKW_73a}                &  $53.3$ & $580$ & $11.041$ & $0.09$ \\
In & $429.8$ &  $920$ & \cite{AB_JCT80a,HDHGKW_73a}                            &  $74.2$ & $600$ &  $9.040$ & $0.16$ \\
Hg & $234.3$ &  $630$ & \cite{SMC_HTHP75a,HDHGKW_73a}                          &  $52.5$ & $260$ &  $8.284$ & $0.04$ \\
Sn & $505.1$ & $1173$ & \cite{GM_SPA66a,B_83a,HDHGKW_73a}                      &  $73.7$ & $640$ &  $9.567$ & $0.15$ \\
Ga & $302.9$ &  $773$ & \cite{KHF_BBPC70a,HDHGKW_73a}                          &  $99.6$ & $360$ &  $7.148$ & $0.07$
\end{tabular}
\end{ruledtabular}
\end{table*}

The second volume correction is negligible until the first volume
correction reaches a magnitude around $0.3Nk_B$.  The second volume
correction is calculated for Hg, since the data are sufficient for
this purpose \cite{SMC_HTHP75a}.  For the alkali metals, the second volume
correction is estimated from experimental data in the vicinity of the
melt curve \cite{MNS_IPC77a}.  For the remaining five liquids, the highest
temperature $T_h$ is such that the second volume correction can be
neglected.

\begin{figure}[h]
\scalebox{0.3}{\includegraphics*{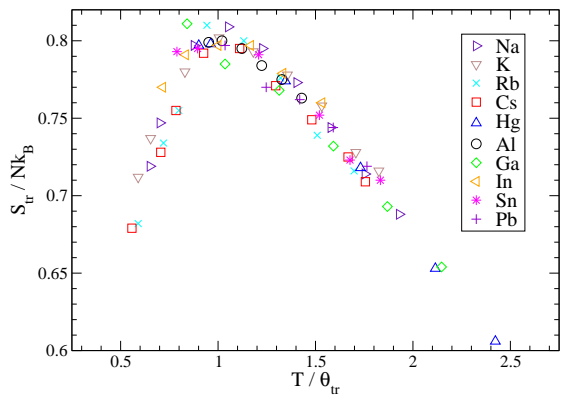}}
\caption{(Color online) Transit entropy for ten liquids calculated from the data
referenced in Table~\ref{maintab}.  Within small scatter, all lie on a
single curve as a function of the scaled temperature, where $\theta_{tr}$ is a
material parameter.  The scatter is small compared to the mean
experimental error (Table~\ref{maintab}).}
\label{Strdata}
\end{figure}

The analysis will now be confined to the volume $V_m$, and the
corresponding notation will be suppressed.  The first step is to find
$S_{tr}$ from $S_{expt}$, using Eq.~(\ref{Sdef}) in the form
\begin{equation}
S_{tr}(T) = S_{expt}(T) - S_{vib}(T) - S_{el}(T).
\label{calcStr}
\end{equation}
$S_{el}(T)$ is calculated from free electron theory in the leading
Sommerfeld expansion.  This is sufficiently accurate for the liquids
studied here, because $S_{el}(T) < 0.02S_{expt}(T)$ throughout the
analysis.  $S_{vib}(T)$ is calculated from Eq.~(\ref{classicalSvib})
with an initial choice for $\theta_0$.  At this point, eight curves of
$S_{tr}$ vs.\ $T$ have clear maxima, with the exceptions being Pb and Ga.
For the eight, the temperature at the maximum is denoted $\theta_{tr}$, and we
graph $S_{tr}(T)$ vs.\ $T/\theta_{tr}$.  The curves look like they will fall
on a single line if they are shifted by various constants in $S_{tr}$.
This is done by varying $\theta_0$, since a change in $\theta_0$
changes $S_{tr}$ by a constant [Eqs.~(\ref{classicalSvib}) and
(\ref{calcStr})].  To bring the curves together, a common value for
the maximum of $S_{tr}$ is needed.  We choose $0.8Nk_B$, the universal
entropy constant of the original V-T formulation.  All ten liquids,
including Pb and Ga, can be shifted to lie on a single smooth curve,
as shown in Fig.~\ref{Strdata}.

In this analysis, the actual data for $S_{tr}(T)$ for each liquid is a
set of points with small scatter.  Therefore, placing the ten data
sets on a single curve is not a precise operation.  But the scatter is
so small that this imprecision is negligible.  The fitted values of
$\theta_0$ and $\theta_{tr}$ are listed in Table~\ref{maintab}.  The total
nuclear motion entropy is
\begin{equation}
S_{nuc}(T) = S_{vib}(T)+S_{tr}(T).
\label{Snucdef}
\end{equation}
To show its magnitude, $S_{nuc}(T_m)$ is also listed in Table~\ref{maintab}.

Aside from experimental data, the information used in the analysis is
Eqs.~(\ref{Sdef})-(\ref{Theta0def}), plus the constraint that all the
curves must have a common maximum of $0.8Nk_B$ at $T=\theta_{tr}$.
For each liquid, this information plus the parameters $\theta_0$ and
$\theta_{tr}$ constitute a fit of the experimental entropy data at
volume $V_m$.  However, we do not expect a precise common maximum to
hold for all monatomic liquids, nor do we expect this maximum to be
volume independent.  For a more quantitative analysis of this issue,
we turn next to the entropy of melting.

\section{Testing V-T theory}
\label{testingVT}

\subsection{Theory for the entropy of melting}
\label{theoryforSmelt}

The experimental information relating to the universality of the
transit entropy is the entropy of melting.  To use this information we
need to express the entropy of melting theory in terms of the entropy
formulation of Sec.~\ref{Sanal}.  To keep the notation simple we shall
continue with our standard notation for the liquid, and use a
superscript $c$ to denote the crystal.  Hence $S$ and $S^c$ are the
entropy of the liquid and crystal respectively, and $V_m$ is the
liquid volume at the melting temperature $T_m$. The measured
constant-pressure entropy of melting, corrected so that both crystal
and liquid have the same volume $V_m$, is
\begin{equation}
\Delta S(V_m,T_m) = S(V_m,T_m) - S^c(V_m,T_m).
\label{DeltaSdef}
\end{equation}
For all the normal melting elements for which accurate experimental
data are available to determine this quantity, the mean and standard
deviations are \cite{W_02a,W_PRSL91a,del_Tl}
\begin{equation}
\Delta S(V_m,T_m) = (0.80 \pm 0.10) Nk_B.
\label{DeltaSval}
\end{equation}
The distribution is essentially the same when $\Delta S$ is evaluated
at the crystal volume $V_m^c$, so the volume dependence is weak.
Among the elements for which accurate experimental data are available,
six do not belong to this distribution and are called
\emph{anomalous}.  Their values of $\Delta S$ in units of $Nk_B$ are
$1.48$ (Sn), $2.37$ (Ga), $2.68$ (Sb), $2.62$ (Bi), $3.76$ (Si), and
$3.85$ (Ge).  The anomalous $\Delta S$ are not only larger than the
normal value, they are \emph{much} larger.  As a fiducial for
elemental metals, the liquid entropy at melt is $S_m \approx 10Nk_B$.
In comparison, experimental error is $\approx 0.5\%$, the width of the
normal $\Delta S$ distribution is very small at $\approx 1\%$, and the
anomalous $\Delta S$ is very large at $10-30\%$.

To rationalize these experimental results, we shall write the V-T
equation for $\Delta S(V_m,T_m)$.  The crystal entropy, with classical
harmonic vibrations, is
\begin{equation}
S^c(V,T) = 3Nk_B\left\{\ln\left[T/\theta_0^c(V)\right] + 1\right\} 
         + S_{el}^c(V,T).
\label{Scr}
\end{equation}
Then with Eqs.~(\ref{Sdef}) and (\ref{classicalSvib}),
\begin{eqnarray}
\Delta S(V_m,T_m) & = & 3Nk_B\ln\left[\theta_0^c(V_m)/\theta_0(V_m)\right] \nonumber \\
                  & & + S_{tr}(V_m,T_m) + \Delta S_{el}(V_m,T_m).
\label{DeltaStheory}
\end{eqnarray}
In normal melting, there is no significant change in the electronic
structure, so that the internuclear forces are nearly the same in
liquid and crystal, and so is the electronic density of states.  Hence
$\theta_0^c(V_m) \approx \theta_0(V_m)$ and $\Delta S_{el}(V_m,T_m)
\approx 0$, so that from Eq.~(\ref{DeltaStheory}), $\Delta S(V_m,T_m)
\approx S_{tr}(V_m,T_m)$.  It then follows from Eq.~(\ref{DeltaSval})
that
\begin{equation}
S_{tr}(V_m,T_m) \approx (0.80 \pm 0.10) Nk_B
\label{Str=0.8}
\end{equation}
for normal melting elements.

In contrast, anomalous melting \emph{is} accompanied by a change in
the electronic structure.  Si and Ge melt from covalent crystal to
metallic liquid (\cite{GCG_69a}, Chap.~3), while Sb and Bi melt from
semimetal crystal to metallic liquid (\cite{F_72a}, p.~81).  For Sn
and Ga, the electronic structure change upon melting becomes apparent
through compression.  Si, Ge, Sb, and Bi all have a triple point on
the melt curve at modest compression \cite{Y_76a}, and so do Sn and Ga
\cite{Y_76a,JKK_PR63a}.  The interpretation is that compression drives
a relative shifting of electronic bands, the shift being continuous
with compression for the liquid, but being concentrated at the
crystal-crystal transition in the solid.  Hence the melting is
anomalous in the vicinity of a triple point.  The classic example is
Cs \cite{JNM_PR67a,YA_JPSJ70a,MPB_JPF74a}, where melting is normal at
ambient pressure, but becomes anomalous under compression when the triple
point is approached (\cite{W_02a}, Fig.~26.5).

Because of the change in electronic structure, the internuclear forces
and electronic density of states are significantly different between
liquid and crystal.  Hence in addition to the normal $S_{tr} \approx
0.8Nk_B$ on the right hand side of Eq.~(\ref{DeltaStheory}), the terms
in nuclear vibration and electronic excitation are both important.
These terms have positive sum because melting is entropy driven, and
the term in $\ln(\theta_0^c/\theta_0)$ is usually dominant.

Let us apply these results to Fig.~\ref{Strdata}.  Equation
(\ref{Str=0.8}) holds for normal melting, and since $S_{tr}(V_m,T)$
changes little between $T_m$ and $\theta_{tr}$, according to
Fig.~\ref{Strdata}, $S_{tr}(V_m,\theta_{tr}) \approx 0.8Nk_B$ for normal
melting elements.  But this is a purely liquid quantity, independent
of the melting process, so this relation should be valid for monatomic
liquids in general.  Let us denote this common maximum $\chi(V)$, a
function of volume.  The above argument suggests
\begin{equation}
\chi(V_m) \approx 0.8Nk_B.
\label{chival}
\end{equation}
In Fig.~\ref{Strdata}, the approximation is taken to be an equality.
In principle, $\chi(V)$ is a material parameter, but as the following
test shows, we are not yet able to resolve specific material
dependence.

\subsection{Testing through \textit{ab initio} calculations}
\label{testthroughabinit}

The test reported here is the comparison of $\theta_0$ from \textit{ab
initio} calculations with results from the analysis of experiment for
Na and Cu.  The technique \cite{BPCDWHL_08a,HBPLDCW_un09a} calculates
the electronic ground state by DFT (using the VASP code \cite{VASP})
for a system of $N$ atoms in a cubic cell, with periodic boundary
conditions on the nuclear positions.  The system is quenched to a
structure, where the frequencies and eigenvectors of the normal
vibrational modes are calculated, and $\theta_0$ is evaluated from
Eq.~(\ref{Theta0def}).  The eigenvalues (mass times squared
frequencies) are always positive, except for the three expressing
translation, which are zero to numerical accuracy.  The structures
found are numerically dominated by random ones.  Each calculation is
done at the density of the liquid at melt, and $\theta_0$(DFT) is
listed in Table~\ref{DFTtab}.

\begin{table}
\caption{DFT calculations of $\theta_0$ for the liquid at $V_m$ \cite{BPCDWHL_08a}, compared with the same quantity determined from experimental entropy data.}
\label{DFTtab}
\begin{ruledtabular}
\begin{tabular}{cccc}
Liquid & $\rho$ (g/cm$^3$) & $\theta_0$(DFT) (K) & $\theta_0$(expt) (K) \\
\hline
Na & $0.935$ & $98.4\pm3.0$ &  $97.6\pm2.3$ \\
Cu & $8.00$  & $171\pm5$     & $171.4\pm5.4$
\end{tabular}
\end{ruledtabular}
\end{table}

Table~\ref{DFTtab} also lists $\theta_0$(expt), which is from
Table~\ref{maintab} for Na.  For Cu, with insufficient data for the
analysis of Sec.~\ref{Sanal}, $\theta_0$(expt) is estimated
separately.  The estimation procedure is general and is outlined for
Cu in Appendix \ref{TtrforCu}.  The agreement between theory and
experiment in Table~\ref{DFTtab} is certainly better than we should
expect.

The application of DFT to liquid dynamics research is currently making
notable progress.  From \textit{ab initio} MD, melting properties of
Si have been calculated by Sugino and Car \cite{SC_PRL95a}, and Wang et
al.\ \cite{WSC_PRL05a} calculated the carbon phase diagram.  This work
points to the possibility of an \textit{ab initio} test of
Eq.~(\ref{DeltaStheory}) for anomalous melting.  High pressure melting
curves have been calculated for Pb by Cricchio et al.\
\cite{CBBPA_PRB06a} and also for Ta by Taioli et al.\
\cite{TCGA_PRB07a}.  In each case the shape of $T_m(P)$ suggests
normal melting, and this again can be tested by calculating the
quantities in Eq.~(\ref{DeltaStheory}).  Kresse's summary of DFT
calculations of the static structure factor and pair distribution
function for group IIIB-VIB elements shows better results for metallic
than nonmetallic liquids \cite{K_JNCS02a}.  While the metallic liquids
should be well described by random valleys, the molecular character of
As, Se, and Te, possessing strong and weak bonds, poses a challenge
regarding the underlying potential energy surface.  A similar
challenge is posed by Ge, whose \textit{ab initio} static and dynamic
structure factors compare well with experiment as shown by Chai et
al.\ \cite{CSHK_PRB03a}, and whose primarily metallic liquid appears
to have some tetrahedral coordination in its fluctuation spectrum at
$T_m$.

The example of highly compressed Na has attracted much attention.  
It was predicted from theory by Neaton and Ashcroft \cite{NA_PRL01a} 
that under compression crystalline Na will transform to low symmetry 
structures that include semimetallic behavior, and tend ultimately to 
semiconducting.  Experiments by Hanfland et al.\ \cite{HLS_PRB02a} and 
Syassen \cite{S_02a} confirmed the structural changes to $120$ GPa, 
and work by Gregoryanz et al.\ \cite{GDSHM_PRL05a} revealed a change 
from normal to anomalous melting.  This was in turn confirmed by 
\textit{ab initio} MD calculations by Raty et al.\ \cite{RSB_N07a}, which 
also showed that the liquid undergoes electronic structure changes 
analogous to those in the solid.  We notice that the anomalous melting 
regime in Na involves $2s$ and $2p$ electrons entering the valence, 
and is related theoretically to the sequence of anomalous melting elements 
Sn, Ga, Sb, and Bi mentioned in Sec.~\ref{theoryforSmelt}.

A novel theoretical technique uses Monte Carlo perturbation theory to
make accurate first principles calculations of the liquid free energy
at arbitrary temperatures \cite{GL_AIP07a,G_JCP08a}.  This technique,
as well as \textit{ab initio} MD, will make possible more accurate
tests of V-T theory than we can obtain through analysis of
experimental data, as in the present study.

\section{Discussion and conclusions}
\label{concl}

In this work, we are able to improve the formulation of the
transit contribution to thermodynamic functions, and to carry out an
\textit{ab initio} test of the V-T theory of liquid entropy.  The
broad implications of these results will be discussed.

\subsection{Transit contribution to thermodynamics}

In the original formulation, transits are accounted for through the
multiplicative factor $\exp(N\ln w)$ in the partition function, where
$\ln w = 0.8$ is calibrated from entropy of melting data.  Hence the
transit free energy is $-Nk_BT\ln w$, the transit entropy is the
universal constant $Nk_B\ln w$, still a good approximation, but the
transit internal energy is zero.  Now, in Sec.~\ref{Sanal}, the
experimental high-temperature entropy data are analyzed, and
Fig.~\ref{Strdata} is found for $S_{tr}(V,T)$ at constant $V$.  In principle
one can integrate the constant-volume relation $dU = T dS$ to find
$U_{tr}(V,T)$ up to a volume-dependent constant of integration.  We
then have the transit free energy $F_{tr}(V,T)$ to replace our
original $-Nk_BT\ln w$ in the liquid free energy.  This accomplishes
the goal set out in the Introduction, and provides in principle the
transit contribution to every thermodynamic function.

The shape of the transit entropy curve in Fig.~\ref{Strdata} can be
understood from qualitative properties of transits.  At very low
temperatures ($T << T_m$) the system freezes into a single random
valley and becomes an amorphous solid.  The motion is entirely
vibrational, there are no transits, and $S_{tr}=0$, as is the
corresponding $U_{tr}$.  Upon warming, transits begin at a given
temperature and both $S_{tr}$ and $U_{tr}$ increase from zero.  This
behavior is seen in MD simulations of supercooled liquid Na, where the
mean potential energy increases from its pure vibrational value of
$(3/2)Nk_BT$ at around $140$ K (\cite{WC_PRE99a}, Fig.~4).  This is
confirmed as a transit effect by observing that self diffusion
increases from zero at approximately the same temperature
(\cite{WC_PRE99a}, Fig.~10).  With increasing temperature, $S_{tr}$
and $U_{tr}$ saturate and have zero slope at the common temperature
$\theta_{tr}$.  The reason for the saturation of $S_{tr}$ and
$U_{tr}$, and their subsequent decrease with increasing $T$, is the
truncation of the random valley potential surface at the intervalley
intersections.  A model for part of this decrease, the boundary
effect, has been applied to the high-temperature specific heat of Hg
\cite{W_PRE98b}.  Now we have more information than specific heat
data, since Fig.~\ref{Strdata} shows both the low-$T$ increase and
high-$T$ decrease of the transit entropy.  Figure \ref{Strdata} explains
the previously disorganized behavior of liquid specific heat curves
(\cite{W_PRE98b}, Fig.~2).  Figure \ref{Strdata} will be helpful in
modeling the statistical mechanics of transits, and especially in
modeling the transit free energy.

In the past we have used $T_m$ as a scaling temperature for liquid
properties.  This is not satisfactory in principle, because $T_m$
depends partially on properties of the crystal.  Not surprisingly,
$T_m$ utterly fails as a liquid scaling temperature for anomalous
melting elements (\cite{W_PRSL91a}, Figs.~3 and 4).  We now have a
two-component theory for the nuclear motion entropy, where each
component has its independent scaling temperature, $\theta_0(V)$ for
$S_{vib}$ and $\theta_{tr}(V)$ for $S_{tr}$.  This is a purely liquid
theory, with no parametric dependence on crystal properties.

\subsection{\textit{Ab initio} testing of V-T theory}

At the time of the original formulation \cite{W_PRE97b}, no potential energy
property of any random valley had been calculated.  Indeed, the
existence of the random and symmetric classes of valleys, and their
contrasting potential energy properties, was only hypothesized in that
first paper.  At the time, we adopted the approximation $\theta_0(V_m)
\approx \theta_0^c(V_m)$ for normal melting elements, since
$\theta_0^c$ was available from force-constant models calibrated to
experimental dispersion curves \cite{SD_81a}.  This is still a respectable
approximation, probably accurate to $3-4\%$ on average for normal
melting elements (but not for anomalous melting elements; see
\cite{CW_PRE04a}, Tables~I and II).  But now, with \textit{ab initio} values
of $\theta_0(V)$, the correct theoretical $S_{vib}(V,T)$ can be
calculated from Eq.~(\ref{classicalSvib}) and compared with 
$S_{vib}(V,T)$ extracted from experiment.  This is essentially the
comparison made in Sec.~\ref{testthroughabinit}.  But that comparison
goes much deeper than a casual glance would suggest.  That test
provides the following extensive support of V-T theory.
\begin{enumerate}
\item The test verifies the original hypothesis that the random
      valleys are numerically dominant and hence account for the entire
      statistical mechanics as $N \rightarrow \infty$, and that the random
      valleys are uniform in their potential properties so that a single
      example is sufficient as $N \rightarrow \infty$.  The verification
      results from the fact that a single random valley is used for the
      calculation of $\theta_0$ for each liquid, while the experiment
      samples enormous numbers of valleys of all types.
\item The test verifies that vibrational motion in a harmonically
      extended random valley is the correct theory for that part of the
      experimental entropy which is identified with $S_{vib}$.  This is
      because that part of the experimental entropy is reproduced by
      \textit{ab initio} evaluation of Eq.~(\ref{classicalSvib}) over
      the entire temperature range of the available experimental data for
      each liquid tested.
\item The test is consistent with $\chi(V_m) = 0.8Nk_B$ for Na and Cu.  
      The expected material dependence of $\chi(V_m)$ is a refinement 
      remaining for future work.
\end{enumerate}
V-T theory is unique in that it offers a Hamiltonian theory capable of
unifying equilibrium and nonequilibrium theories of liquid dynamics.
Further testing will help to develop a robust theory.  Techniques
currently being applied in \textit{ab initio} simulations can be
employed to advantage in the development of V-T theory.

\appendix

\section{Error estimates}
\label{Strerrors}

Error in $S_{expt}(V_m,T)$ arises from experimental error in the
high-$T$ entropy data, and the experimental error in the volume
correction.  We estimate the mean of each error over the range from
$T_m$ to $T_h$, and add their magnitudes to obtain the total mean
experimental error $\langle \delta S_{expt} \rangle$, as listed in
Table~\ref{maintab}.  Much of this error is already present in the
data for $S_{expt}(V_m, T_m)$.  The remaining error is $T$-dependent
and has an average around zero for each liquid.  This error is large
enough to cause a noticeable error in the shape of Fig.~\ref{Strdata}.

The relative experimental error in $\theta_0$ is 
$\delta\theta_0\textrm{(expt)} / \theta_0$, and is almost entirely due
to $\langle \delta S_{expt}\rangle$.  From Eq.~(\ref{classicalSvib})
it follows that 
\begin{equation} 
\frac{\delta\theta_0\textrm{(expt)}}{\theta_0} = \frac{\langle \delta S_{expt}\rangle}{3Nk_B}.
\label{deltatheta}
\end{equation}
This gives a range of $1-5\%$ for the experimental error in $\theta_0$
in Table~\ref{maintab}.  The error in $\theta_{tr}$ is entirely due to
the $T$-dependent error, because $\theta_{tr}$ depends on the shape of
the curve and not on its magnitude.  The error in $\theta_{tr}$ in
Table~\ref{maintab} can reach $10\%$.

Moments of the frequency distribution can be calculated from DFT to an
accuracy of $1\%$ for elemental crystals.  These crystal calculations
are done in the infinite lattice model \cite{BH_54a}, where an
arbitrary number of Brillouin-zone $\bm{k}$-points is possible.  In
contrast, the liquid system has only $3N$ normal models for an
$N$-atom system.  Because of the small system size ($N=150$) for our
DFT calculations, the present error in $\theta_0\textrm(DFT)$ is
allowed as $3\%$.

\section{Estimation of scaling temperatures for \NoCaseChange{Cu}}
\label{TtrforCu}

The slope of $S_{tr}(T_m)$ is $C_{tr}(T_m)$, the transit contribution
to the specific heat, which can be found from data as follows.  The
experimental specific heat at constant volume is $C_{expt}$, while the
vibrational contribution in classical statistical mechanics is
$C_{vib} = 3Nk_B$, so that $C_{tr} = C_{expt} - 3Nk_B - C_{el}$.
Matching this quantity, with error estimates, to the slope of a smooth
curve fitted to Fig.~\ref{Strdata} yields $T_m/\theta_{tr}$ in the
range $0.8-1.3$ for Cu at $T_m$.  Then Fig.~\ref{Strdata} implies that 
$S_{tr}(V_m,T_m)$ is $(0.78-0.80)Nk_B$.  From experimental entropy
data for Cu, $S_{nuc}(T_m) = 10.00 Nk_B$, so that $S_{vib}(T_m)=
(9.21\pm0.01)Nk_B$, giving $\theta_0 = (171.4 \pm 0.6)$ K.  For Cu,
$\langle \delta S_{expt} \rangle$ is just the estimated experimental entropy
error at $T_m$, namely $0.086\,Nk_B$.

This method becomes inaccurate when $|C_{tr}|$ is large.  For a
compilation of $C_{vib} + C_{tr}$ for elemental liquids at melt, see
\cite{W_PRE97b}.

\acknowledgments{For many helpful discussions and insights we thank
Brad Clements, Giulia De Lorenzi-Venneri, Carl Greeff, Erik
Holmstr\"{o}m, and Travis Peery.  This work was supported by the
U.~S.~DOE under Contract No.~DE-AC52-06NA25396.}

\bibliography{VTref_2Apr2009} 

\begin{thebibliography}{56}
\expandafter\ifx\csname natexlab\endcsname\relax\def\natexlab#1{#1}\fi
\expandafter\ifx\csname bibnamefont\endcsname\relax
  \def\bibnamefont#1{#1}\fi
\expandafter\ifx\csname bibfnamefont\endcsname\relax
  \def\bibfnamefont#1{#1}\fi
\expandafter\ifx\csname citenamefont\endcsname\relax
  \def\citenamefont#1{#1}\fi
\expandafter\ifx\csname url\endcsname\relax
  \def\url#1{\texttt{#1}}\fi
\expandafter\ifx\csname urlprefix\endcsname\relax\def\urlprefix{URL }\fi
\providecommand{\bibinfo}[2]{#2}
\providecommand{\eprint}[2][]{\url{#2}}

\bibitem[{\citenamefont{Boltzmann}(1872)}]{B_WB72a}
\bibinfo{author}{\bibfnamefont{L.}~\bibnamefont{Boltzmann}},
  \bibinfo{journal}{Wien.\ Ber.} \textbf{\bibinfo{volume}{66}},
  \bibinfo{pages}{275} (\bibinfo{year}{1872}).

\bibitem[{\citenamefont{Born and Huang}(1954)}]{BH_54a}
\bibinfo{author}{\bibfnamefont{M.}~\bibnamefont{Born}} \bibnamefont{and}
  \bibinfo{author}{\bibfnamefont{K.}~\bibnamefont{Huang}},
  \emph{\bibinfo{title}{Dynamical Theory of Crystal Lattices}}
  (\bibinfo{publisher}{Clarendon Press, Oxford}, \bibinfo{year}{1954}).

\bibitem[{\citenamefont{Born and von Karmen}(1912)}]{BvK_PZ12a}
\bibinfo{author}{\bibfnamefont{M.}~\bibnamefont{Born}} \bibnamefont{and}
  \bibinfo{author}{\bibfnamefont{Th.}~\bibnamefont{von Karmen}},
  \bibinfo{journal}{Phys.\ Z.\ } \textbf{\bibinfo{volume}{13}},
  \bibinfo{pages}{297} (\bibinfo{year}{1912}).

\bibitem[{\citenamefont{Wallace}(1997)}]{W_PRE97b}
\bibinfo{author}{\bibfnamefont{D.~C.} \bibnamefont{Wallace}},
  \bibinfo{journal}{Phys.\ Rev.\ E} \textbf{\bibinfo{volume}{56}},
  \bibinfo{pages}{4179} (\bibinfo{year}{1997}).

\bibitem{macro_equiv} Macroscopic equivalence applies to one liquid at
one density.  Statistical properties of random valleys vary with
density, and vary from one liquid to another.

\bibitem[{\citenamefont{Wallace}(2002)}]{W_02a}
\bibinfo{author}{\bibfnamefont{D.~C.} \bibnamefont{Wallace}},
  \emph{\bibinfo{title}{Statistical Physics of Crystals and Liquids}}
  (\bibinfo{publisher}{World Scientific, New Jersey}, \bibinfo{year}{2002}).

\bibitem[{\citenamefont{{De~Lorenzi-Venneri}
  et~al.}(2008)\citenamefont{{De~Lorenzi-Venneri}, Chisolm, and
  Wallace}}]{DCW_PRE08a}
\bibinfo{author}{\bibfnamefont{G.}~\bibnamefont{{De~Lorenzi-Venneri}}},
  \bibinfo{author}{\bibfnamefont{E.~D.} \bibnamefont{Chisolm}},
  \bibnamefont{and} \bibinfo{author}{\bibfnamefont{D.~C.}
  \bibnamefont{Wallace}}, \bibinfo{journal}{Phys.\ Rev.\ E}
  \textbf{\bibinfo{volume}{78}}, \bibinfo{pages}{041205}
  (\bibinfo{year}{2008}).

\bibitem[{\citenamefont{Harrison}(1966)}]{H_66a}
\bibinfo{author}{\bibfnamefont{W.~A.} \bibnamefont{Harrison}},
  \emph{\bibinfo{title}{Pseudopotentials in the Theory of Metals}}
  (\bibinfo{publisher}{W.\ A.\ Benjamin, New York}, \bibinfo{year}{1966}).

\bibitem[{\citenamefont{Ashcroft}(1966)}]{A_PL66a}
\bibinfo{author}{\bibfnamefont{N.~W.} \bibnamefont{Ashcroft}},
  \bibinfo{journal}{Phys.\ Lett.} \textbf{\bibinfo{volume}{23}},
  \bibinfo{pages}{48} (\bibinfo{year}{1966}).

\bibitem[{\citenamefont{Faber}(1972)}]{F_72a}
\bibinfo{author}{\bibfnamefont{T.~E.} \bibnamefont{Faber}},
  \emph{\bibinfo{title}{Theory of Liquid Metals}}
  (\bibinfo{publisher}{Cambridge University Press, Cambridge},
  \bibinfo{year}{1972}).

\bibitem[{\citenamefont{Ashcroft and Stroud}(1978)}]{AS_SSP78a}
\bibinfo{author}{\bibfnamefont{N.~W.} \bibnamefont{Ashcroft}} \bibnamefont{and}
  \bibinfo{author}{\bibfnamefont{D.}~\bibnamefont{Stroud}},
  \bibinfo{journal}{Solid State Phys.\ } \textbf{\bibinfo{volume}{33}},
  \bibinfo{pages}{1} (\bibinfo{year}{1978}).

\bibitem[{\citenamefont{March}(1990)}]{M_90a}
\bibinfo{author}{\bibfnamefont{N.~H.} \bibnamefont{March}},
  \emph{\bibinfo{title}{Liquid Metals: Concepts and Theory}}
  (\bibinfo{publisher}{Cambridge University Press, Cambridge, England},
  \bibinfo{year}{1990}).

\bibitem[{\citenamefont{Wallace and Clements}(1999)}]{WC_PRE99a}
\bibinfo{author}{\bibfnamefont{D.~C.} \bibnamefont{Wallace}} \bibnamefont{and}
  \bibinfo{author}{\bibfnamefont{B.~E.} \bibnamefont{Clements}},
  \bibinfo{journal}{Phys.\ Rev.\ E} \textbf{\bibinfo{volume}{59}},
  \bibinfo{pages}{2942} (\bibinfo{year}{1999}).

\bibitem[{\citenamefont{Clements and Wallace}(1999)}]{CW_PRE99a}
\bibinfo{author}{\bibfnamefont{B.~E.} \bibnamefont{Clements}} \bibnamefont{and}
  \bibinfo{author}{\bibfnamefont{D.~C.} \bibnamefont{Wallace}},
  \bibinfo{journal}{Phys.\ Rev.\ E} \textbf{\bibinfo{volume}{59}},
  \bibinfo{pages}{2955} (\bibinfo{year}{1999}).

\bibitem[{\citenamefont{Chisolm et~al.}(2001)\citenamefont{Chisolm, Clements,
  and Wallace}}]{CCW_PRE01a}
\bibinfo{author}{\bibfnamefont{E.~D.} \bibnamefont{Chisolm}},
  \bibinfo{author}{\bibfnamefont{B.~E.} \bibnamefont{Clements}},
  \bibnamefont{and} \bibinfo{author}{\bibfnamefont{D.~C.}
  \bibnamefont{Wallace}}, \bibinfo{journal}{Phys.\ Rev.\ E}
  \textbf{\bibinfo{volume}{63}}, \bibinfo{pages}{031204}
  (\bibinfo{year}{2001}).

\bibitem[{\citenamefont{{De~Lorenzi-Venneri} and Wallace}(2007)}]{DW_PRE07a}
\bibinfo{author}{\bibfnamefont{G.}~\bibnamefont{{De~Lorenzi-Venneri}}}
  \bibnamefont{and} \bibinfo{author}{\bibfnamefont{D.~C.}
  \bibnamefont{Wallace}}, \bibinfo{journal}{Phys.\ Rev.\ E}
  \textbf{\bibinfo{volume}{76}}, \bibinfo{pages}{041203}
  (\bibinfo{year}{2007}).

\bibitem[{\citenamefont{Bock et~al.}(2008)\citenamefont{Bock, Peery, Chisolm,
  {De~Lorenzi-Venneri}, Wallace, Holmstr{\"o}m, and Liz{\'a}rraga}}]{BPCDWHL_08a}
\bibinfo{author}{\bibfnamefont{N.}~\bibnamefont{Bock}},
  \bibinfo{author}{\bibfnamefont{T.}~\bibnamefont{Peery}},
  \bibinfo{author}{\bibfnamefont{E.~D.} \bibnamefont{Chisolm}},
  \bibinfo{author}{\bibfnamefont{G.}~\bibnamefont{{De~Lorenzi-Venneri}}},
  \bibinfo{author}{\bibfnamefont{D.~C.} \bibnamefont{Wallace}},
  \bibinfo{author}{\bibfnamefont{E.}~\bibnamefont{Holmstr{\"o}m}},
  \bibnamefont{and}
  \bibinfo{author}{\bibfnamefont{R.}~\bibnamefont{Liz{\'a}rraga}},
  \bibinfo{journal}{Bull.\ Am.\ Phys.\ Soc.\ } \textbf{\bibinfo{volume}{53}}(2),
  \bibinfo{pages}{J9:00004} (\bibinfo{year}{2008}).

\bibitem[{\citenamefont{Holmstr{\"o}m et~al.}()\citenamefont{Holmstr{\"o}m,
  Bock, Peery, Liz{\'a}rraga, {De Lorenzi-Venneri}, Chisolm, and
  Wallace}}]{HBPLDCW_un09a}
\bibinfo{author}{\bibfnamefont{E.}~\bibnamefont{Holmstr{\"o}m}},
  \bibinfo{author}{\bibfnamefont{N.}~\bibnamefont{Bock}},
  \bibinfo{author}{\bibfnamefont{T.}~\bibnamefont{Peery}},
  \bibinfo{author}{\bibfnamefont{R.}~\bibnamefont{Liz{\'a}rraga}},
  \bibinfo{author}{\bibfnamefont{G.}~\bibnamefont{{De Lorenzi-Venneri}}},
  \bibinfo{author}{\bibfnamefont{E.~D.}~\bibnamefont{Chisolm}},
  \bibnamefont{and}
  \bibinfo{author}{\bibfnamefont{D.~C.}~\bibnamefont{Wallace}},
  \bibinfo{howpublished}{unpublished}.

\bibitem[{\citenamefont{Wallace}(1991)}]{W_PRSL91a}
\bibinfo{author}{\bibfnamefont{D.~C.} \bibnamefont{Wallace}},
  \bibinfo{journal}{Proc.\ R.\ Soc.\ London, Ser.\ A} \textbf{\bibinfo{volume}{433}},
  \bibinfo{pages}{615} (\bibinfo{year}{1991}).

\bibitem[{\citenamefont{Gol'tsova}(1966)}]{G_HT66a}
\bibinfo{author}{\bibfnamefont{E.~I.} \bibnamefont{Gol'tsova}},
  \bibinfo{journal}{High Temp.} \textbf{\bibinfo{volume}{4}},
  \bibinfo{pages}{348} (\bibinfo{year}{1966}).

\bibitem[{\citenamefont{Hultgren et~al.}(1973)\citenamefont{Hultgren, Desai,
  Hawkins, Gleiser, Kelley, and Wagman}}]{HDHGKW_73a}
\bibinfo{author}{\bibfnamefont{R.}~\bibnamefont{Hultgren}},
  \bibinfo{author}{\bibfnamefont{P.~D.} \bibnamefont{Desai}},
  \bibinfo{author}{\bibfnamefont{D.~T.} \bibnamefont{Hawkins}},
  \bibinfo{author}{\bibfnamefont{M.}~\bibnamefont{Gleiser}},
  \bibinfo{author}{\bibfnamefont{K.~K.} \bibnamefont{Kelley}},
  \bibnamefont{and} \bibinfo{author}{\bibfnamefont{D.~D.}
  \bibnamefont{Wagman}}, \emph{\bibinfo{title}{Selected Values of the
  Thermodynamic Properties of the Elements}} (\bibinfo{publisher}{ASM, Metals
  Park, OH}, \bibinfo{year}{1973}).

\bibitem[{\citenamefont{Shaw and Caldwell}(1985)}]{SC_PRB85a}
\bibinfo{author}{\bibfnamefont{G.~H.} \bibnamefont{Shaw}} \bibnamefont{and}
  \bibinfo{author}{\bibfnamefont{D.~A.} \bibnamefont{Caldwell}},
  \bibinfo{journal}{Phys.\ Rev.\ B} \textbf{\bibinfo{volume}{32}},
  \bibinfo{pages}{7937} (\bibinfo{year}{1985}).

\bibitem[{\citenamefont{Trelin et~al.}(1966)\citenamefont{Trelin, Vasil'ev,
  Proskurin, and Tsyganova}}]{YTVPT_HT66a}
\bibinfo{author}{\bibfnamefont{Yu.~S.} \bibnamefont{Trelin}},
  \bibinfo{author}{\bibfnamefont{I.~N.} \bibnamefont{Vasil'ev}},
  \bibinfo{author}{\bibfnamefont{V.~B.} \bibnamefont{Proskurin}},
  \bibnamefont{and} \bibinfo{author}{\bibfnamefont{T.~A.}
  \bibnamefont{Tsyganova}}, \bibinfo{journal}{High Temp.}
  \textbf{\bibinfo{volume}{4}}, \bibinfo{pages}{352} (\bibinfo{year}{1966}).

\bibitem[{\citenamefont{{Chase, Jr.} et~al.}(1985)\citenamefont{{Chase, Jr.},
  Davies, {Downey, Jr.}, Frurip, McDonald, and Syverud}}]{CDDFMS_JPCRD85a}
\bibinfo{author}{\bibfnamefont{M.~W.} \bibnamefont{{Chase, Jr.}}},
  \bibinfo{author}{\bibfnamefont{C.~A.} \bibnamefont{Davies}},
  \bibinfo{author}{\bibfnamefont{J.~R.} \bibnamefont{{Downey, Jr.}}},
  \bibinfo{author}{\bibfnamefont{D.~J.} \bibnamefont{Frurip}},
  \bibinfo{author}{\bibfnamefont{R.~A.} \bibnamefont{McDonald}},
  \bibnamefont{and} \bibinfo{author}{\bibfnamefont{A.~N.}
  \bibnamefont{Syverud}}, \bibinfo{journal}{J.\ Phys.\ Chem.\ Ref.\ Data Suppl.\ No. 1}
  \textbf{\bibinfo{volume}{14}}, 61 (\bibinfo{year}{1985}).

\bibitem[{\citenamefont{Shpil'rain et~al.}(1985)\citenamefont{Shpil'rain,
  Yakimovich, Fomin, Skovorodjko, and Mozgovoi}}]{SYFSM_85a}
\bibinfo{author}{\bibfnamefont{E.~E.} \bibnamefont{Shpil'rain}},
  \bibinfo{author}{\bibfnamefont{K.~A.} \bibnamefont{Yakimovich}},
  \bibinfo{author}{\bibfnamefont{V.~A.} \bibnamefont{Fomin}},
  \bibinfo{author}{\bibfnamefont{S.~N.} \bibnamefont{Skovorodjko}},
  \bibnamefont{and} \bibinfo{author}{\bibfnamefont{A.~G.}
  \bibnamefont{Mozgovoi}}, in \emph{\bibinfo{booktitle}{Handbook of
  Thermodynamic and Transport Properties of Alkali Metals}}, edited by
  \bibinfo{editor}{\bibfnamefont{R.~W.} \bibnamefont{Ohse}}
  (\bibinfo{publisher}{Blackwell, London}, \bibinfo{year}{1985}), p.
  \bibinfo{pages}{435}.

\bibitem[{\citenamefont{Vargaftik et~al.}(1977)\citenamefont{Vargaftik,
  Kozhevnikov, Stepanov, Alekseev, and Ryzhkov}}]{VKSAR_77a}
\bibinfo{author}{\bibfnamefont{N.~B.} \bibnamefont{Vargaftik}},
  \bibinfo{author}{\bibfnamefont{V.~F.} \bibnamefont{Kozhevnikov}},
  \bibinfo{author}{\bibfnamefont{V.~G.} \bibnamefont{Stepanov}},
  \bibinfo{author}{\bibfnamefont{V.~A.} \bibnamefont{Alekseev}},
  \bibnamefont{and} \bibinfo{author}{\bibfnamefont{Y.~F.}
  \bibnamefont{Ryzhkov}}, in \emph{\bibinfo{booktitle}{Seventh Symposium on
  Thermophysical Properties}}, edited by
  \bibinfo{editor}{\bibfnamefont{A.}~\bibnamefont{Cezairliyan}}
  (\bibinfo{publisher}{ASME, New York}, \bibinfo{year}{1977}), p.
  \bibinfo{pages}{926}.

\bibitem[{\citenamefont{Brandes}(1983)}]{B_83a}
\bibinfo{author}{\bibfnamefont{E.~A.} \bibnamefont{Brandes}},
  \emph{\bibinfo{title}{Smithells Metals Reference Book}}
  (\bibinfo{publisher}{Butterworths, London}, \bibinfo{year}{1983}).

\bibitem[{\citenamefont{Miller}(1952)}]{M_52a}
\bibinfo{author}{\bibfnamefont{R.~R.} \bibnamefont{Miller}}, in
  \emph{\bibinfo{booktitle}{Liquid Metals Handboook}}, 2nd ed., edited by
  \bibinfo{editor}{\bibfnamefont{R.~N.} \bibnamefont{Lyon}}
  (\bibinfo{publisher}{U.\ S.\ Government Printing Office, Washington, D.\ C.},
  \bibinfo{year}{1952}), p.~\bibinfo{pages}{38}.

\bibitem[{\citenamefont{Tsu et~al.}(1982)\citenamefont{Tsu, Suenaga, Takano,
  and Shiraishi}}]{TSTS_TJIM82a}
\bibinfo{author}{\bibfnamefont{Y.}~\bibnamefont{Tsu}},
  \bibinfo{author}{\bibfnamefont{H.}~\bibnamefont{Suenaga}},
  \bibinfo{author}{\bibfnamefont{K.}~\bibnamefont{Takano}}, \bibnamefont{and}
  \bibinfo{author}{\bibfnamefont{Y.}~\bibnamefont{Shiraishi}},
  \bibinfo{journal}{Trans.\ Japn.\ Inst.\ Metals} \textbf{\bibinfo{volume}{23}},
  \bibinfo{pages}{1} (\bibinfo{year}{1982}).

\bibitem[{\citenamefont{Gitis and Mikhailov}(1966)}]{GM_SPA66a}
\bibinfo{author}{\bibfnamefont{M.~B.} \bibnamefont{Gitis}} \bibnamefont{and}
  \bibinfo{author}{\bibfnamefont{I.~G.} \bibnamefont{Mikhailov}},
  \bibinfo{journal}{Sov.\ Phys.\ Acoust.\ } \textbf{\bibinfo{volume}{11}},
  \bibinfo{pages}{372} (\bibinfo{year}{1966}).

\bibitem[{\citenamefont{Almond and Blairs}(1980)}]{AB_JCT80a}
\bibinfo{author}{\bibfnamefont{D.~P.} \bibnamefont{Almond}} \bibnamefont{and}
  \bibinfo{author}{\bibfnamefont{S.}~\bibnamefont{Blairs}},
  \bibinfo{journal}{J.\ Chem.\ Thermodyn.\ } \textbf{\bibinfo{volume}{12}},
  \bibinfo{pages}{1105} (\bibinfo{year}{1980}).

\bibitem[{\citenamefont{Spetzler et~al.}(1975)\citenamefont{Spetzler, Myer, and
  Chan}}]{SMC_HTHP75a}
\bibinfo{author}{\bibfnamefont{H.~A.} \bibnamefont{Spetzler}},
  \bibinfo{author}{\bibfnamefont{M.~D.} \bibnamefont{Myer}}, \bibnamefont{and}
  \bibinfo{author}{\bibfnamefont{T.}~\bibnamefont{Chan}},
  \bibinfo{journal}{High Temp. - High Press.} \textbf{\bibinfo{volume}{7}},
  \bibinfo{pages}{481} (\bibinfo{year}{1975}).

\bibitem[{\citenamefont{K{\"o}ster et~al.}(1970)\citenamefont{K{\"o}ster,
  Hensel, and Franck}}]{KHF_BBPC70a}
\bibinfo{author}{\bibfnamefont{H.}~\bibnamefont{K{\"o}ster}},
  \bibinfo{author}{\bibfnamefont{F.}~\bibnamefont{Hensel}}, \bibnamefont{and}
  \bibinfo{author}{\bibfnamefont{E.~U.} \bibnamefont{Franck}},
  \bibinfo{journal}{Ber.\ Bunsenges.\ Phys.\ Chem.}
  \textbf{\bibinfo{volume}{74}}, \bibinfo{pages}{43} (\bibinfo{year}{1970}).

\bibitem[{\citenamefont{Makarenko et~al.}(1977)\citenamefont{Makarenko,
  Nikolaenko, and Stishov}}]{MNS_IPC77a}
\bibinfo{author}{\bibfnamefont{I.~N.} \bibnamefont{Makarenko}},
  \bibinfo{author}{\bibfnamefont{A.~M.} \bibnamefont{Nikolaenko}},
  \bibnamefont{and} \bibinfo{author}{\bibfnamefont{S.~M.}
  \bibnamefont{Stishov}}, \bibinfo{journal}{Inst.\ Phys.\ Conf.\ Ser.\ }{\bf 30},
  \bibinfo{pages}{79} (\bibinfo{year}{1977}).

\bibitem{del_Tl} In the years between \cite{W_PRSL91a} and
\cite{W_02a}, we realized that Tl should be removed from the analysis
because available crystal data are for hcp, while melting proceeds
from bcc.  This changes the average $\Delta S / Nk_B$ insignificantly,
from $0.79$ to $0.80$.

\bibitem[{\citenamefont{Glazov et~al.}(1969)\citenamefont{Glazov, Chizhevskaya,
  and Glagoleva}}]{GCG_69a}
\bibinfo{author}{\bibfnamefont{V.~M.} \bibnamefont{Glazov}},
  \bibinfo{author}{\bibfnamefont{S.~N.} \bibnamefont{Chizhevskaya}},
  \bibnamefont{and} \bibinfo{author}{\bibfnamefont{N.~N.}
  \bibnamefont{Glagoleva}}, \emph{\bibinfo{title}{Liquid Semiconductors}}
  (\bibinfo{publisher}{Plenum, New York}, \bibinfo{year}{1969}).

\bibitem[{\citenamefont{Young}(1976)}]{Y_76a}
\bibinfo{author}{\bibfnamefont{D.~A.} \bibnamefont{Young}},
  \emph{\bibinfo{title}{Phase Diagrams of the Elements}}
  (\bibinfo{publisher}{University of California Press, Berkeley, CA},
  \bibinfo{year}{1976}).

\bibitem[{\citenamefont{Jayaraman et~al.}(1963)\citenamefont{Jayaraman,
  {Klement, Jr.}, and Kennedy}}]{JKK_PR63a}
\bibinfo{author}{\bibfnamefont{A.}~\bibnamefont{Jayaraman}},
  \bibinfo{author}{\bibfnamefont{W.}~\bibnamefont{{Klement, Jr.}}},
  \bibnamefont{and} \bibinfo{author}{\bibfnamefont{G.~C.}
  \bibnamefont{Kennedy}}, \bibinfo{journal}{Phys.\ Rev.}
  \textbf{\bibinfo{volume}{130}}, \bibinfo{pages}{540} (\bibinfo{year}{1963}).

\bibitem[{\citenamefont{Jayaraman et~al.}(1967)\citenamefont{Jayaraman, Newton,
  and McDonough}}]{JNM_PR67a}
\bibinfo{author}{\bibfnamefont{A.}~\bibnamefont{Jayaraman}},
  \bibinfo{author}{\bibfnamefont{R.~C.} \bibnamefont{Newton}},
  \bibnamefont{and} \bibinfo{author}{\bibfnamefont{J.~M.}
  \bibnamefont{McDonough}}, \bibinfo{journal}{Phys.\ Rev.}
  \textbf{\bibinfo{volume}{159}}, \bibinfo{pages}{527} (\bibinfo{year}{1967}).

\bibitem[{\citenamefont{Yamashita and Asano}(1970)}]{YA_JPSJ70a}
\bibinfo{author}{\bibfnamefont{J.}~\bibnamefont{Yamashita}} \bibnamefont{and}
  \bibinfo{author}{\bibfnamefont{S.}~\bibnamefont{Asano}},
  \bibinfo{journal}{J.\ Phys.\ Soc.\ Jpn.\ } \textbf{\bibinfo{volume}{29}},
  \bibinfo{pages}{264} (\bibinfo{year}{1970}).

\bibitem[{\citenamefont{McWhan et~al.}(1974)\citenamefont{McWhan, Parisot, and
  Bloch}}]{MPB_JPF74a}
\bibinfo{author}{\bibfnamefont{D.~B.} \bibnamefont{McWhan}},
  \bibinfo{author}{\bibfnamefont{G.}~\bibnamefont{Parisot}}, \bibnamefont{and}
  \bibinfo{author}{\bibfnamefont{D.}~\bibnamefont{Bloch}},
  \bibinfo{journal}{J.\ Phys.\ F} \textbf{\bibinfo{volume}{4}},
  \bibinfo{pages}{L69} (\bibinfo{year}{1974}).

\bibitem[{VAS()}]{VASP}
http://cms.mpi.univie.ac.at/vasp/

\bibitem[{\citenamefont{Sugino and Car}(1995)}]{SC_PRL95a}
\bibinfo{author}{\bibfnamefont{O.}~\bibnamefont{Sugino}} \bibnamefont{and}
  \bibinfo{author}{\bibfnamefont{R.}~\bibnamefont{Car}},
  \bibinfo{journal}{Phys.\ Rev.\ Lett.} \textbf{\bibinfo{volume}{74}},
  \bibinfo{pages}{1823} (\bibinfo{year}{1995}).

\bibitem[{\citenamefont{Wang et~al.}(2005)\citenamefont{Wang, Scandolo, and
  Car}}]{WSC_PRL05a}
\bibinfo{author}{\bibfnamefont{X.}~\bibnamefont{Wang}},
  \bibinfo{author}{\bibfnamefont{S.}~\bibnamefont{Scandolo}}, \bibnamefont{and}
  \bibinfo{author}{\bibfnamefont{R.}~\bibnamefont{Car}},
  \bibinfo{journal}{Phys.\ Rev.\ Lett.} \textbf{\bibinfo{volume}{95}},
  \bibinfo{pages}{185701} (\bibinfo{year}{2005}).

\bibitem[{\citenamefont{Cricchio et~al.}(2006)\citenamefont{Cricchio,
  Belonoshko, Burakovsky, Preston, and Ahuja}}]{CBBPA_PRB06a}
\bibinfo{author}{\bibfnamefont{F.}~\bibnamefont{Cricchio}},
  \bibinfo{author}{\bibfnamefont{A.~B.} \bibnamefont{Belonoshko}},
  \bibinfo{author}{\bibfnamefont{L.}~\bibnamefont{Burakovsky}},
  \bibinfo{author}{\bibfnamefont{D.~L.} \bibnamefont{Preston}},
  \bibnamefont{and} \bibinfo{author}{\bibfnamefont{R.}~\bibnamefont{Ahuja}},
  \bibinfo{journal}{Phys.\ Rev.\ B} \textbf{\bibinfo{volume}{73}},
  \bibinfo{pages}{140103(R)} (\bibinfo{year}{2006}).

\bibitem[{\citenamefont{Taioli et~al.}(2007)\citenamefont{Taioli, Cazorla,
  Gillan, and Alf{\`e}}}]{TCGA_PRB07a}
\bibinfo{author}{\bibfnamefont{S.}~\bibnamefont{Taioli}},
  \bibinfo{author}{\bibfnamefont{C.}~\bibnamefont{Cazorla}},
  \bibinfo{author}{\bibfnamefont{M.~J.} \bibnamefont{Gillan}},
  \bibnamefont{and} \bibinfo{author}{\bibfnamefont{D.}~\bibnamefont{Alf{\`e}}},
  \bibinfo{journal}{Phys.\ Rev.\ B} \textbf{\bibinfo{volume}{75}},
  \bibinfo{pages}{214103} (\bibinfo{year}{2007}).

\bibitem[{\citenamefont{Kresse}(2002)}]{K_JNCS02a}
\bibinfo{author}{\bibfnamefont{G.}~\bibnamefont{Kresse}}, \bibinfo{journal}{J.\
  Non-Cryst.\ Solids} \textbf{\bibinfo{volume}{312-314}}, \bibinfo{pages}{52}
  (\bibinfo{year}{2002}).

\bibitem[{\citenamefont{Chai et~al.}(2003)\citenamefont{Chai, Stroud, Hafner,
  and Kresse}}]{CSHK_PRB03a}
\bibinfo{author}{\bibfnamefont{J.-D.} \bibnamefont{Chai}},
  \bibinfo{author}{\bibfnamefont{D.}~\bibnamefont{Stroud}},
  \bibinfo{author}{\bibfnamefont{J.}~\bibnamefont{Hafner}}, \bibnamefont{and}
  \bibinfo{author}{\bibfnamefont{G.}~\bibnamefont{Kresse}},
  \bibinfo{journal}{Phys.\ Rev.\ B} \textbf{\bibinfo{volume}{67}},
  \bibinfo{pages}{104205} (\bibinfo{year}{2003}).

\bibitem[{\citenamefont{Neaton and Ashcroft}(2001)}]{NA_PRL01a}
\bibinfo{author}{\bibfnamefont{J.~B.} \bibnamefont{Neaton}} \bibnamefont{and}
  \bibinfo{author}{\bibfnamefont{N.~W.} \bibnamefont{Ashcroft}},
  \bibinfo{journal}{Phys.\ Rev.\ Lett.} \textbf{\bibinfo{volume}{86}},
  \bibinfo{pages}{2830} (\bibinfo{year}{2001}).

\bibitem[{\citenamefont{Hanfland et~al.}(2002)\citenamefont{Hanfland, Loa, and
  Syassen}}]{HLS_PRB02a}
\bibinfo{author}{\bibfnamefont{M.}~\bibnamefont{Hanfland}},
  \bibinfo{author}{\bibfnamefont{I.}~\bibnamefont{Loa}}, \bibnamefont{and}
  \bibinfo{author}{\bibfnamefont{K.}~\bibnamefont{Syassen}},
  \bibinfo{journal}{Phys.\ Rev.\ B} \textbf{\bibinfo{volume}{65}},
  \bibinfo{pages}{184109} (\bibinfo{year}{2002}).

\bibitem[{\citenamefont{Syassen}(2002)}]{S_02a}
\bibinfo{author}{\bibfnamefont{K.}~\bibnamefont{Syassen}}, in
  \emph{\bibinfo{booktitle}{High Pressure Phenomena}}, edited by
  \bibinfo{editor}{\bibfnamefont{R.~J.} \bibnamefont{Hemley}},
  \bibinfo{editor}{\bibfnamefont{G.}~\bibnamefont{Chiarotti}},
  \bibinfo{editor}{\bibfnamefont{M.}~\bibnamefont{Bernasconi}},
  \bibnamefont{and} \bibinfo{editor}{\bibfnamefont{L.}~\bibnamefont{Ulivi}}
  (\bibinfo{publisher}{IOS, Amsterdam}, \bibinfo{year}{2002}), p.
  \bibinfo{pages}{251}.

\bibitem[{\citenamefont{Gregoryanz et~al.}(2005)\citenamefont{Gregoryanz,
  Degtyareva, Somayazulu, Hemley, and Mao}}]{GDSHM_PRL05a}
\bibinfo{author}{\bibfnamefont{E.}~\bibnamefont{Gregoryanz}},
  \bibinfo{author}{\bibfnamefont{O.}~\bibnamefont{Degtyareva}},
  \bibinfo{author}{\bibfnamefont{M.}~\bibnamefont{Somayazulu}},
  \bibinfo{author}{\bibfnamefont{R.~J.} \bibnamefont{Hemley}},
  \bibnamefont{and} \bibinfo{author}{\bibfnamefont{H.~K.}~\bibnamefont{Mao}},
  \bibinfo{journal}{Phys.\ Rev.\ Lett.} \textbf{\bibinfo{volume}{94}},
  \bibinfo{pages}{185502} (\bibinfo{year}{2005}).

\bibitem[{\citenamefont{Raty et~al.}(2007)\citenamefont{Raty, Schwegler, and
  Bonev}}]{RSB_N07a}
\bibinfo{author}{\bibfnamefont{J.-Y.} \bibnamefont{Raty}},
  \bibinfo{author}{\bibfnamefont{E.}~\bibnamefont{Schwegler}},
  \bibnamefont{and} \bibinfo{author}{\bibfnamefont{S.~A.} \bibnamefont{Bonev}},
  \bibinfo{journal}{Nature (London)} \textbf{\bibinfo{volume}{449}},
  \bibinfo{pages}{448} (\bibinfo{year}{2007}).

\bibitem[{\citenamefont{Greeff and Liz{\'a}rraga}(2007)}]{GL_AIP07a}
\bibinfo{author}{\bibfnamefont{C.~W.} \bibnamefont{Greeff}} \bibnamefont{and}
  \bibinfo{author}{\bibfnamefont{R.}~\bibnamefont{Liz{\'a}rraga}},
  {\it Shock Compression of Condensed Matter - 2007},
  \bibinfo{journal}{AIP Conf.\ Proc.\ No.\ 955} (AIP, New York, 2007), p.\ 43.

\bibitem[{\citenamefont{Greeff}(2008)}]{G_JCP08a}
\bibinfo{author}{\bibfnamefont{C.~W.} \bibnamefont{Greeff}},
  \bibinfo{journal}{J.\ Chem.\ Phys.} \textbf{\bibinfo{volume}{128}},
  \bibinfo{pages}{184104} (\bibinfo{year}{2008}).

\bibitem[{\citenamefont{Wallace}(1998)}]{W_PRE98b}
\bibinfo{author}{\bibfnamefont{D.~C.} \bibnamefont{Wallace}},
  \bibinfo{journal}{Phys.\ Rev.\ E} \textbf{\bibinfo{volume}{57}},
  \bibinfo{pages}{1717} (\bibinfo{year}{1998}).

\bibitem[{\citenamefont{Schober and Dederichs}(1981)}]{SD_81a}
\bibinfo{author}{\bibfnamefont{H.}~\bibnamefont{Schober}} \bibnamefont{and}
  \bibinfo{author}{\bibfnamefont{P.~H.} \bibnamefont{Dederichs}}, in
  \bibinfo{booktitle}{Landoldt-Bornstein New Series Vol.\ 13a}, edited by
  \bibinfo{editor}{\bibfnamefont{K.-H.} \bibnamefont{Hellwege}}
  (\bibinfo{publisher}{Springer, Berlin}, \bibinfo{year}{1981}).

\bibitem[{\citenamefont{Chisolm and Wallace}(2004)}]{CW_PRE04a}
\bibinfo{author}{\bibfnamefont{E.~D.} \bibnamefont{Chisolm}} \bibnamefont{and}
  \bibinfo{author}{\bibfnamefont{D.~C.} \bibnamefont{Wallace}},
  \bibinfo{journal}{Phys.\ Rev.\ E} \textbf{\bibinfo{volume}{69}},
  \bibinfo{pages}{031204} (\bibinfo{year}{2004}).

\end{thebibliography}

\end{document}